\documentclass[prb,preprint]{revtex4-1} 
\usepackage{amsmath}  % needed for \tfrac, \bmatrix, etc.
\usepackage{amsfonts} % needed for bold Greek, Fraktur, and blackboard bold
\usepackage{graphicx} % needed for figures
\usepackage{epstopdf}
\usepackage{esint}

\begin{document}

% Be sure to use the \title, \author, \affiliation, and \abstract macros
% to format your title page.  Don't use lower-level macros to  manually
% adjust the fonts and centering.

\title{Draft: Volume Integral of the Pressure Gradient and Archimede's Principle}
% In a long title you can use \\ to force a line break at a certain location.

\author{Galen T. Pickett}
\email{Galen.Pickett@csulb.edu} % optional
\affiliation{Department of Physics and Astronomy, California State University Long Beach, 1240 Bellflower Blvd., Long Beach, CA 90840, USA} 

% See the REVTeX documentation for more examples of author and affiliation lists.

\date{\today}

\begin{abstract}
The theorems of vector analysis (divergence theorem, etc.) are typically first applied in the undergraduate physics curriculum in the context of the electromagnetic field and the differential forms of Maxwell's equations.
However, these tools are analyzed in depth several courses later in the junior-senior level.  I discuss here a ``bridge" problem, using the language of vector calculus in a mechanics setting to understand Archimedes' principle as a consequence of 
hydrostatic equilibrium and
the superposition of the external forces.  
It is my hope that this treatment will help students better integrate and understand understand these and similar vector analysis results in contexts beyond electromagnetism.
\end{abstract}
% AJP requires an abstract for all regular article submissions.
% Abstracts are optional for submissions to the "Notes and Discussions" section.

\maketitle % title page is now complete

\section{Introduction}
There is a qualitative transition in the sophistication of the mathematical tools required to describe mechanical systems of a few free bodies to those dominated by 
the behavior of continuous fields.
Second-order ordinary differential equations are indispensable in the study of point-system mechanics, particularly in a first year course, and many curricula are 
organized around first understanding the properties of these equations (particularly in the case in which the equation of motion has the form 
$\ddot{x} = const$).
In a first course in electromagnetism, the fundamental quantities to understand are the distributed electromagnetic fields, $\vec{E},\vec{B}$, with
dynamics given by Maxwell's Equations using the language of partial differential equations with appropriate boundary conditions.  Even very good companion works can give the impression to undergraduates that the ideas of ``divergence, gradient, and curl'' have their primary (perhaps only) role in describing electricity and magnetism. \cite{dgcandallthat}
Thus, beginning students get the impression that these mathematical tools are somehow wedded to the separate subjects of the physics canon, rather than being independent, widely applicable tools.

My aim here is to present a mechanics problem which is usually dealt with in the algebra-based mechanics curriculum and recast it using the mathematical language 
students expect in electricity and magnetism. I have two aims.  First, I will connect Archimedes' principle (usually developed as a ``just-so" statement of how a ``buoyant force" works) to real forces acting on real bodies.  
The ideas of buoyancy, density, and volume are known areas of difficulty for science teaching candidates, K-12 students, and even quite advanced students majoring
in physics. \cite{loverude1,loverude2}
As we shall see, the buoyant force is merely a consequence of the superposition of many forces acting
on the outer edge of an object, and thus is no more mysterious than any other force in mechanics.
Secondly, I wish to give an undergraduate a context in which vector analysis has a definite application in the most familiar, concrete physics a student knows, namely mechanics.

Thus, the integration of these mathematical tools and the ``net forces" ideas in this nexus of problems and placing that integrated
understanding firmly within the grasp
of an intermediate-level student of physics is the goal.
It is not my intention to merely display yet another derivation of the principle, although I will certainly do that.  
Rather, it is my intention to integrate this basic idea into our curriculum in a fundamental way.
If this work is used to merely guide a new set of derivations in a lecture hall, its purpose will be unfulfilled.
There is a growing consensus across the
academy that the value of integrating methods and content, and creating a unified, flexible intellectual framework supporting lifelong learning is
the eventual goal of a baccalaureate course of study --- even in professional and STEM fields, perhaps {\it particularly so}.  
Too often a university education can be experienced as a list of topics or courses logically disconnected from each other.  
The work of integrating the education into a seamless whole can be left to a student, and may or may not occur. \cite{aacu_integrate}
Physics is particularly blessed in this regard, in that our discipline celebrates the unification of disparate phenomena in a way that few disciplines have a history of doing.  
There are a number of excellent introductory texts that take this integrated point of view as opposed to a historical development.  The ``right"
ideas are developed at the beginning of study, and the special cases amenable to analysis (exact or approximate) are labeled as such.\cite{mandi}

In the physics program at CSU Long Beach where I am a faculty member, we have for a number of years been intentional with this very issue, stemming from our activity as a PhysTEC supported site 
in 2010.\cite{phystec}
One element of this program was to develop a Learning Assistant Program on the Colorado model, along with a course in physics pedagogy \cite{phys390}
we use to train physics majors to teach in our lower-level laboratories as Learning Assistants. \cite{la_model}
A part of that course, and the key to explaining to physics majors why this low-level material is so hard to learn (and teach) is discussing
 the excellent article by Reddish on mental models in physics learning. \cite{reddish}
One of the most striking features of this treatment is the realization that mental structures of physics in a student's head are incomplete, disorganized, and even for very well-trained students (and faculty) there exists ``holes'' in even very basic concepts.

In taking this course, students are required to be present in a science tutoring center, and as is often the case, have duties to help students with courses they themselves had never taken.  Algebra-based mechanics and K-6 level physical sciences are never taken by physics students, and we assume that advanced students (with 
advanced mathematical and physics tools) should be able to handle the more elementary ideas in those ``precursor'' courses.

It is a relatively simple problem that stumped my stable of learning assistants that prompted the development below --- a problem using Archimedes' principle.
It is my experience that anything involving fluids is inordinately tricky to explain.  
The basic functioning of a centrifuge can be understood at an engineering level and qualitatively well enough for the devices to work correctly in the hands of research-caliber scientists and engineers.  
Yet, the explanation in terms of hydrostatic equilibrium in a non-inertial frame is subtle with tendrils reaching up into very sophisticated mechanics.\cite{taylor,goldstein} 
Of the many  treatments of the principle (starting with a good exposition of Archimedes' words themselves \cite{whatsaid} and even quite interesting
derivations centering on energy ideas \cite{apfrome}) not many cement the principle into the basic foundation of mechanics 
many students understand and can already use quite well.
Even in a static case, in which Archimedes' principle and buoyancy should be sufficient to solve many problems, my group of junior level physics students ``knew'' Archimedes' principle, but were unable to apply it.  
And, embarrassingly enough, likewise for myself.  
Only when I had put the problem into the language of vector analysis did Archimedes' principle actually make sense to me (in the sense that I had connected it to my own ideas of mechanics). 
The point of this paper is to elucidate the conditions that lead me to ``reconstruct'' a theorem of vector calculus Archimedes himself must have discovered: \cite{joke}
\begin{equation}
\varoiint d \vec{A} p = \iiint dv \vec{\nabla} p
\label{at}
\end{equation}
which some will recognize as the three-dimensional version of the ``mean-value theorem,'' but which I will simply regard as ``Archimedes' Theorem" ---
a fitting companion to Gauss' law.

I will first describe the elementary problem that revealed to us all that we had a buoyancy gap in our understanding.
Then I will describe hydrostatic equilibrium and the new vector identity, and in the discussion describe some applications and ``clicker'' style questions.

\section{Elementary Problem}
The problem that send my physics major Learning Assistants off into a cycle of recursive algebra is a relatively straightforward problem encountered in the algebra-based mechanics sequence.  Here is a variation:

{\it A hot air balloon of volume $V$ is in equilibrium surrounded by air of density $\rho_a$.  At $t=0$ ballast of mass $m_o$ is released from the balloon, and it is observed to accelerate upwards with an acceleration $a$.  What is the final mass of the balloon (in terms of $m_o, \rho_a, g, V$)?}

These were students, by and large, who had just taken our junior-level course in mechanics, or had completed a course in introductory modern physics and differential equations, so there was no issue with their preparation.  
Yet, they reported that they spent an inordinate amount of time circling around the problem without coming to a resolution, much to the consternation of the the first-year students who were hoping to get some help on the problem in our tutoring center.

What my students knew, and what they had connected to the stock of problems they had personally solved, were two different things.  It was only the presence of our Master Teacher in Residence, Ms. Kathryn Beck of Bolsa Grande High School that saved our collective bacon by remarking that the buoyant force is just $\rho_a V g$ both before the ballast had been dropped and after.\cite{tir}  Given that this force is the same before and after leads immediately to the observation that
\begin{equation}
M = m_o g / a
\end{equation}
In short, my students had ``heard'' of Archimedes' principle (as I had myself), but in never having had to use it in a physical context (this sort of problem has fallen out of our ``calculus''-based mechanics curriculum).
They were quite confident in their use of free body diagrams, but the nature of the buoyant force was not connected from the beginning to the end of the problem.

And, once I had the benefit of Ms. Beck's observation (she was clearly very, very familiar with this sort of problem, having taught it many years in many guises and having
charged generations of students to construct and predict the displacement of cardboard boats) I was left with a conundrum.  There is in reality 
no {\bf buoyant} force, there is only the effect of the surrounding air pressure on the balloon.  As I had drilled into my own mind in the course of working through my own Ph.D. thesis \cite{pickett}, only
gradients of the pressure field give forces.
The presence of a pressure gradient is the {\bf only} way the air can affect the balloon, not through an unsystematic 
``watch me pull a rabbit out of a hat'' buoyant force.

\section{Pressure}
And, here is how I connect the buoyant force and the presence of a non-uniform pressure field.
\begin{figure}[h!]
\centering
\includegraphics[scale=0.5]{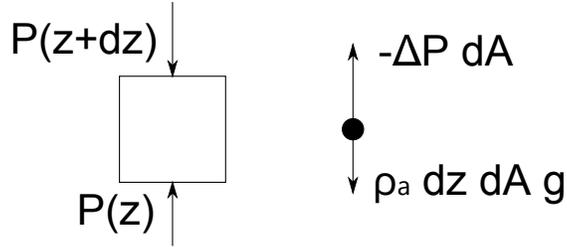}
\caption{A small cube of the ambient air at density $\rho$ is subjected to an inward force on its upper face, an upward force on its lower face, and its weight. 
 An abstract ``net force" or ``free body" diagram is at right.}
\label{schematic}
\end{figure}
The air around the balloon is in hydrostatic equilibrium.  Thus, a small cube of air with edge lengths  $dx, dy, dz$  is in equilibrium when:
\begin{equation}
(\rho_a dx dy dz) g = P(x,y,z) dx dy - P(x,y,z+dz) dx dy
\end{equation}
Here, the pressure on the top face of the box produces a downward force, and the pressure on the bottom face produces an upward force.
Rearranging a bit gives a differential equation for the pressure:
\begin{equation}
\rho_a g = -  \partial_z P
\label{pgrad}
\end{equation}
with the familiar solution:
\begin{equation}
P = P_o - \rho_a g z,
\label{hydro}
\end{equation}
where $P_o$ is the pressure at $z=0$.  This treatment is elementary enough, certainly not worth writing a manuscript to share.

\begin{figure}[h!]
\centering
\includegraphics[scale=0.5]{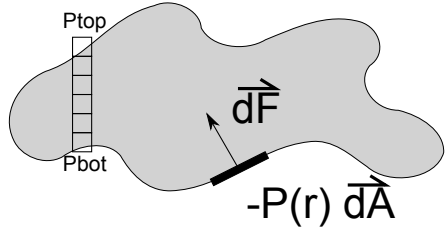}
\caption{An arbitrary shaped object is impinged by pressure forces from the 
surrounding medium.  Given the pressure field, $P(\vec{r})$, our task is to
add up all of the forces resulting from each infinitesimal directed area element,
$d\vec{A}$.  Also shown schematically is a ``box discretization'' of the sort
used  to motivate theorems in vector calculus.\cite{dgcndallthat}}
\label{at_schematic}
\end{figure}

So, now we know the pressure field around the balloon.  What is the microscopic mechanism of transferring force to the balloon from this pressre field?  Again, mirroring the constructions in Ref.~\cite{dgcandallthat} I consider an infinitesimal element of the surface area (with normal directed outward) of the ballon, $d\vec{A}$.
As in Figure~\ref{at_schematic}, the air puts a force on this area element, $d\vec{F} = -P d\vec{A}$. 
Note that there is a pressure field inside the balloon as well, but if we are concerned with the balloon as the system, all such internal forces cancel in pairs --- a distinct advantage of the Matter and Interactions curriculum\cite{mandi} is the clear demarcation between system and surroundings in every phsyical situation.

Thus, the total force from the air is a sum over all of these small air forces:
\begin{equation}
\vec{F} = \varoiint - d\vec{A} P
\label{fair}
\end{equation}
This looks suspiciously like the ``flux'' expression in Gauss' Law, except $\vec{F}_a$ is a vector
quantity and not a scalar formed by the flux of a vector field through a closed surface.

Here is the clever bit, which generates a new (to me) theorem of vector calculus.  Let us dot bot sides of Eq.~\ref{fair} with $\hat{z}$:
\begin{equation}
F_z = \varoiint - d \vec{A} \cdot (P \hat{z}) = 
\iiint dv -\vec{\nabla} ( P \hat{z} ).
\end{equation}
The second equality follows form the divergence theorem applied to the vector field $P \hat{z}$.  
Thus, we have
\begin{equation}
F_z = - \iiint dv \partial_z P = - (-g) \iiint dv \rho_a = g  M_a,
\end{equation}
where the pressure gradient in Eq.~\ref{pgrad} is used in the second equality.

Eureka.

This gives just the $z$ component of the total force, and in this case, as the pressure has no gradients in either the $x$ nor $y$ directions, this is all the analysis can tell us, although if $P$ depends on each of $x,y, z$ nontrivially, we have derived the the full version of Archimedes' theorem,
Eq.~\ref{at}.

\section{Discussion}
The development of Archimedes' Principle as a consequence of hydrostatic equilibrium, Eq.~\ref{hydro} and a generalization of the divergence theorem, 
Eq.~\ref{at} is clearly within the ability of an advanced undergraduate to use, and, in my opinion, relaxes the appearance of the Archimedes' Principle from an {\it ad
hoc} statement and grounds it solidly upon the development of general equilibrium, forces, and superposition of forces.
It is altogether more clear to an advanced student {\it why the principle is true} which allows it to be used in different contexts.

The design of numerous hand-made ``bob'' accelerometers is a case in point.  Here, an object less dense than water is submerged completely, yet tethered to the
bottom of a clear, plastic bottle.  
The bob reacts quickly to an applied acceleration ... accelerate forward, and the bob immediately leans forward, and the magnitude of the
acceleration can be measured by the angle ot the tilt.
I have seen many students fumbling with {\bf why} this should occur.  There is some loose talk of ``water rushing backward, the bob getting out of the way,'' and
yet this is clearly an unsystematic treatment.
Suppose the bottle is accelerated along the $x$-axis.
A pressure field is immediately set up in the fluid (on the timescale it would take a shockwave to traverse the bottle) with a large pressure at the rear
of the bottle, and a smaller pressure at the front.
This component of the gradient of the pressure field has to match the hydrostatic equilibrium condition, Eq.~\ref{hydro}, with an acceleration $\vec{a}$:
\begin{equation}
P = P_o + (\vec{g}-\vec{a}) \cdot \vec{r}.
\end{equation}
Then, an application of Eq.~\ref{at} yields
\begin{equation}
\vec{F}_{fluid} = (\vec{a}-\vec{g}) M_{fluid}.
\end{equation}
Thus, for a quiescent bottle, there is a net pressure force in the direction opposite ``$\vec{g}$'', but given the equivalence principle, we should be able to replace 
$\vec{g}$ with an accelerating noninertial frame with acceleration equal to $-\vec{g}$ with no observable effects.
And, if we moved the whole discussion off into the Internationa Space Station, we would have $\vec{g}=0$, and to get the bob to point away form the bottom of the bottle, we would have to accelerate the bottle in the $\hat{z}$ direction with the magnitude of the terrestrial $g$ to get the same behavior.

A suitable probe for the reasonableness of this whole discussion is the following ``clicker"-style question.
{\it A helium-filled balloon is suspended from a thread one meter from the floor of an enclosed elevator.  At the instant you cut the thread, the cable supporting the
elevator is severed.  While you are hurtling toward certain severe injury, you observe the balloon 
\begin{itemize}
\item{A)} Rise toward the ceiling of the elevator.
\item{B)} Remain one meter from the floor of the elevator.
\item{C)} Fall toward the floor of the elevator.
\end{itemize}
}
The correct answer is B), of course, for a very physical reason that is opaque in the usual reasoning of the principle.  In the freely falling elevator, the pressure field is uniform, and hence the net force from the air on the helium balloon is zero.  The balloon thus has only its weight force acting on it, and it freely falls along with the elevator.
An unsophisticated reading of the problem will assume that the buoyant force is {\bf always} acting, so the balloon will not only rise upward, it will rise upward faster and faster until it hits the roof of the elevator.  A suitable demonstration (in a very special case) is found in dropping a water-filled container with a neutrally buoyant object floating in its midsection.  
If the buoyant force always acts, the ball will remain in equilibrium while the container falls, 
and he top of the container will fall toward the ball.
Yet, in the freely falling system, the water, ball, and container all should share the same acceleration.

And further, we have a solid explanation in terms of real forces in an inertial frame for the functioning of a centrifuge.  For, as the centrifuge spins, there is an 
acceleration field of the form $ - \omega^2 \vec{r}_{\perp}$.
Thus, the equivalent of the hydrostatic condition is (in cylindrical coordinates)
\begin{equation}
\partial_r P = -\omega^2 r \rho_l
\end{equation}
with the analog of Eq.~\ref{at} is then
\begin{equation}
F_{fluid} = omega^2 \rho_l \iiint dv = \omega^2 \rho_l M_l R
\end{equation}
where $R$ radial position of the geometric center of the displaced volume.  This is exactly the position of the center of mass of the fluid that would exists had the
object not been submerged in the fluid.
The dependence on $R$ on $F_{fluid}$ provides an enhancement of the density-discriminating effect, crucial for the operation of a gas-centrifuge.
But again, this is solely a function of adding up the pressure on each surface of an object, and determining the net force.
From there, we can connect to a student's early training in free body diagrams (and the Momentum Principle, in the language of Ref.~\cite{mandi}.

What is going on here is extremely clear from a computational point of view.  As in Figure~\ref{at_schematic}, we attempt to calculate the effect of the pressure forces
on the finite elements in the entire body under consideration.  As the pressures are all equal in the $x$ and $y$ directions, it is just the $z$ direction
that I will concentrate upon here.
Each element of volume has two faces, and the pressures at neighboring surfaces give canceling forces.  It is just the pressure at the top, and the pressure
at the bottom of the overall column that is important, and this gradient is just adding up the increments in the weights of the fluid elements.
What we have here is little more than
\begin{equation}
\Delta P = dx dy  (P_{top} - P_{n-1} + P_{n-1} - P_{n-2} + \cdots P_1 - P_{bot})
\end{equation}
with 
\begin{equation}
(P_{i} - P_{i-1}) dx dy = (dx dy dz) \rho g.
\end{equation}
A student asked to {\bf compute the effect of the pressure gradient} will find Eq~\ref{at} as a trivial restatement of a massive internal cancellation
of terms.  
Exactly this sort of exercise is well within the capability of students completing a computation-enriched curriculum.\cite{mandi,taylor,goldstein}

\section{Conclusion}
Here, I have related an episode revealing a ``hole'' in the preparation of undergraduate students in physics at CSU Long Beach 
(and perhaps in many more institutions beside), and proposed a solution.
The solution brings the mathematics of vector analysis back into the physics curriculum between the hiatus in electrodynamics courses
(just when it would be very useful {\it e.g.} in an introduction to quantum mechanics).  
And, the argument provides a clear insight into how deeply Archimedes himself understood the theorems of vector calculus.\cite{joke}


\begin{thebibliography}{99}
\bibitem{phys390} PHYS 390 ``Exploring Physics Teaching'' is taken by approximately 30 physics majors anually at CSU Long Beach, and counts
as upper-division elective credit in the physics major.
\bibitem{joke} While there is evidence that Archimedes had developed ideas of infinitesimals in various geometric problems, he almost certainly had no 
concept of a vector field or vector calculus.  It is a testament to his powers of insight that his principle has such deep roots in vector analysis.
\end{thebibliography}
\end{document}